\pdfoutput=1 
\documentclass[aps,prl,twocolumn,preprintnumbers,floatfix,reprint]{revtex4-1}
\usepackage{amsmath,amssymb, amsfonts, bbm}
\usepackage{graphics,epsfig,color,subfigure}

\usepackage{mathrsfs, hyperref, natbib}

\newcommand{\be}{\begin{equation}}
\newcommand{\ee}{\end{equation}}

\newcommand{\dslash}{D\!\!\!\!\slash \,\,}

\def\bea{\begin{align}}
\def\ena{\end{align}}

\def\Tr{\mbox{ Tr }}
\def\beqa{\begin{eqnarray}}
\def\enqa{\end{eqnarray}}

\begin{document}

\title{Orbifold equivalence and the sign problem at finite baryon density}

\author{Aleksey Cherman$^1$}
\thanks{Present address: DAMTP, University of Cambridge, Cambridge CB3 0WA, U.K.}
\email{a.cherman@damtp.cam.ac.uk} 
\author{Masanori Hanada$^2$}
\thanks{Present address: Department of Physics, University of Washington, Seattle, WA 98195-1560, USA.}
\email{mhanada@u.washington.edu} 
\author{Daniel Robles-Llana$^2$}
\email{daniel.robles@weizmann.ac.il}
\affiliation{
$^1$ Maryland Center for Fundamental Physics, Department of Physics, University of Maryland,
College Park, Maryland 20742-4111, USA\\
$^2$ Department of Particle Physics and Astrophysics, Weizmann Institute of Science, Rehovot 76100, Israel}

\preprint{WIS/13/10-AUG-DPPA}
\preprint{UMD/DOE-40762-487}

\begin{abstract}
We point out that $SO(2N_{c})$ gauge theory with $N_{f}$ fundamental Dirac fermions does not have a sign problem at finite baryon number chemical potential $\mu_{B}$.   One can thus use lattice Monte Carlo simulations to study this theory at finite density.   The absence of a sign problem in the $SO(2N_{c})$ theory is particularly interesting because a wide class of observables in the $SO(2N_{c})$ theory coincide with observables in QCD in the large $N_{c}$ limit, as we show using the technique of large $N_{c}$ orbifold equivalence.  We argue that the orbifold equivalence between the two theories continues to hold at finite $\mu_{B}$ provided one adds appropriate deformation terms to the $SO(2N_{c})$ theory.  This opens up the prospect of learning about QCD at finite $\mu_{B}$ using lattice studies of the $SO(2N_{c})$ theory.
\end{abstract}

\maketitle
The properties of QCD at high baryon densities have long been a subject of intense interest.  Apart from its intrinsic theoretical appeal, this subject is important in astrophysics, especially in the study of neutron stars.   Because of  asymptotic freedom, the behavior of QCD at asymptotically high chemical potential for baryon number $\mu_{B}$ is well understood theoretically, and QCD becomes a color superconductor as $\mu_{B} \rightarrow \infty$~\cite{Alford:2007xm}.   At more  phenomenologically realistic densities, QCD is strongly coupled, and thus not amenable to controlled analytic treatment.  Lattice Monte Carlo simulation is very useful at $\mu_{B}=0$.  However, it runs into trouble at $\mu_{B}\neq 0$ due to the fermion sign problem: the fermion determinant becomes complex, rendering importance sampling exponentially difficult.  

Over the years, several gauge theories that do not suffer from sign problems at finite density have been explored. The main examples are QCD with an isospin chemical potential\cite{Son:2000xc}, two-color QCD\cite{Dagotto:1986gw,*Hands:1999md,Kogut:1999iv,*Kogut:2000ek}, and adjoint QCD\cite{Kogut:1999iv,*Kogut:2000ek,Hands:2000ei,Auzzi:2006ns,*Bolognesi:2007ut,*Auzzi:2008hu}. However, while interesting, these theories have many qualitative differences from $N_{c}=3$ QCD, such as {\it e.g.} explicitly broken flavor symmetry in the first case.

Here, we propose a path to study QCD at $\mu_{B}\neq 0$ in the large $N_c$ limit using lattice Monte Carlo methods.  Large $N_{c}$ QCD~\cite{Hooft:1973jz,*Witten:1979kh}  gives many insights into nonperturbative strong interactions at zero $\mu_{B}$: it is often a good approximation to our $N_{c}=3$ world.  The extent to which the finite-density large $N_{c}$ world is a good approximation to the $N_{c}=3$ world is discussed in, for instance, Refs.~\cite{Shuster:1999tn,*Park:1999bz,*Frandsen:2005mb,*Buchoff:2009za,*McLerran:2007qj,*Torrieri:2010gz}.  

Our proposal rests on two observations.  The first is that all representations of $SO(2N_{c})$ are real, and as a result $SO(2N_{c})$ gauge theory with $N_{f}$ fundamental Dirac fermions does not have a sign problem at $\mu_{B} \neq 0$.  This alone already makes the theory worth studying, especially because the $SO(2N_{c})$ theory shares a number of qualitative features with $N_{c}\ge 3$ QCD.  For example, it has $2 N_{c}$-valence-quark baryons~\cite{Witten:1983tx}.  

Our second observation is that the connection between the $SO(2N_{c})$ theory and $SU(N_{c})$ QCD is in fact quantitative. We show that $SU(N_c)$ gauge theory with $N_f$ fundamental Dirac fermions ({\it i.e.}, large $N_{c}$ QCD) can be obtained as an orbifold projection of the $SO(2N_c)$ theory.  Large $N_c$ orbifold equivalence \cite{Kachru:1998ys,*Bershadsky:1998cb,*Schmaltz:1998bg,Kovtun:2003hr,*Kovtun:2004bz,*Kovtun:2005kh,Armoni:2003gp} then guarantees that all correlation functions of operators in the ``neutral'' sector (i.e., invariant under the symmetry used for the projection) coincide in both theories to leading order in the $1/N_{c}$ expansion, provided the symmetries used in the projection are not spontaneously broken.  The necessary symmetries are unbroken  at $\mu_{B} = 0$, and thus the $SO(2N_{c})$ theory and large $N_{c}$ QCD have coinciding correlation functions for a broad class of operators.  The equivalence should continue to hold at $\mu_{B} \neq 0$, provided one adds certain deformation terms to the $SO(2N_{c})$ theory which protect the orbifold symmetry, but do not otherwise affect the connection of the theory to large $N_{c}$ QCD.   We show that there exist deformations that protect the orbifold symmetry at least for $\mu_{B} \ll \Lambda_{QCD}$, and likely for larger $\mu_{B}$ as well, all while keeping the theory sign problem free in the chiral limit.  The existence of a sign-problem-free theory equivalent to finite-density large $N_{c}$ QCD is unexpected and quite remarkable.

The orbifold equivalence between the two theories at $\mu_{B}=0$ can be checked using lattice simulations, as can the question of whether the necessary symmetries are protected at large $\mu_{B}$.   If the proposal passes these checks, the $SO(2N_{c})$ gauge theory may be used to perform nonperturbative studies of large $N_c$ QCD at finite density.  

{\it $SO(2N_{c})$ gauge theory. } The $SO(2N_{c})$ 4D gauge theory with $N_{f}$ flavors of Dirac fermions (in Euclidean signature) is
\be 
\label{eq:SOLagrangian}
\mathcal{L_{SO}} 
=
\frac{1}{4 g_{SO}^{2} } \Tr F_{\mu \nu}^2
+ 
\sum_{a =1}^{N_{f}}
\bar{q}_{a} (\gamma^{\mu} D_{\mu} + m_{q}+\mu_{B}\gamma^{4}) q_{a}
\ee
where $F_{\mu \nu}$ is the $SO(2N_{c})$ field strength, $D_{\mu} = \partial_{\mu}+i A_{\mu}$, $q_{a}$ is a Dirac fermion in the fundamental representation of $SO(2N_{c})$, and $m_{q}$ and $\mu_{B}$ are the quark mass and chemical potential. $A_{\mu}= A_{\mu}^{i} t_{i}$, where the $t_{i}$ are the generators of  $SO(2N_{c})$;  we take $\Tr t_{i} t_{j} = \delta_{ij}$.  

When $m_{q} = \mu_{B}=0$, Eq.~\eqref{eq:SOLagrangian} has an $SU(N_{f})_{L}\times SU(N_{f})_{R} \times U(1)_{B} \times U(1)_{A}$ chiral symmetry at the classical level, just like $SU(N_{c})$ QCD.  However, the chiral symmetry of the theory is actually larger than this, since $SO(2N_{c})$ is a real gauge group;
classically it extends to $U(2N_{f})$~\cite{Coleman:1980mx,*Peskin:1980gc} .  $U(1)_A \subset U(2N_{f})$ is anomalous as usual at finite $N_{c}$.  The chiral condensate $\bar{q}{q}$ breaks to $SO(2N_{f}) \supseteq SU(N_{f})_{V} $.  The resulting Nambu-Goldstone bosons (NGBs), with mass $m_{\pi} \sim \sqrt{m_{q}}$, live on $SU(2N_{f})/SO(2N_{f})$.  Some of the NGBs, the ``pions'', are pseudoscalars that couple to $\bar{q}_a \gamma_{5} q_b$, while the others, which we will refer to as baryonic pions, are charged under $U(1)_{B}$.  The baryonic pions are parity even~\cite{Cherman:2011mh} and couple to color-singlet operators of the form $S_{a b}$ and $S_{a b}^{\dag}$, where $S_{a b} = q_{a}^{T} C\gamma^{5}  q_{b}$ and $C=\gamma_{4}\gamma_{2}$ is the charge conjugation matrix satisfying $C\gamma_\mu C^{-1}=-\gamma_\mu^\ast$.
 
There are $N_{f}(N_{f}+1)$ baryonic pions in the theory, and $N_{f}^{2}-1$ pions with no baryon number.    The $SO(2N_{c})$ theory also contains baryon-number-charged cousins of other mesons normally encountered in QCD.  

Now consider turning on $\mu_{B}\neq0$. Since in the chiral limit the baryonic pions are the lightest particles charged under $U(1)_{B}$, once $\mu_{B}\ge m_{\pi}/2$, one would expect the system to undergo a second-order phase transition to a phase with a nonzero density of baryonic pions.  In fact, on general grounds, one expects that the baryonic pions will Bose condense.   In other theories with ``baryonic pions'', namely 2-color QCD and adjoint QCD, explicit chiral perturbation theory ($\chi$PT) calculations show that this does indeed happen~\cite{Kogut:1999iv,*Kogut:2000ek}. We expect the same in the $SO(2N_{c})$ theory~\cite{Cherman:2011mh}.  The baryonic pion condensate breaks $U(1)_{B}\rightarrow \mathbb{Z}_{2}$.   The breaking of $U(1)_{B}$ at $\mu_{B} \geq m_{\pi}/2$ in the $SO(2N_{c})$ gauge theory is in sharp contrast to the way $SU(N_{c})$ QCD behaves, where there are no baryonic pions to be condensed.  We return to this crucial point below, in the context of orbifold projections.

{\it Orbifold projection to $SU(N_{c})$. }   To perform an orbifold projection, one identifies a discrete subgroup of the symmetry group of the ``parent'' theory, which for us is the $SO(2N_{c})$ theory, and sets to zero all of the degrees of freedom in the parent theory that are not invariant under the discrete symmetry.  This gives a ``daughter'' theory, which in this case turns out to be large $N_{c}$ QCD.  The orbifold projection uses a $\mathbb{Z}_{2}$ subgroup of the $SO(2N_{c})\times U(1)_{B}$ symmetry of the $SO(2N_{c})$ theory.   

To define the orbifold projection, take $J\in SO(2N_{c})$ to be given by $J = i\sigma_{2} \otimes 1_{N_{c}}$; $1_{N}$ is an $N \times N$ identity matrix.  [For earlier work on projections from $SO(2N_{c})$ to $SU(N_{c})$, see \cite{Cicuta:1982fu,*Lovelace:1982hz,*Unsal:2006pj}.]  $J$ generates a $\mathbb{Z}_{4}$ subgroup of $SO(2N_{c})$.  Next, let $\omega = e^{i \pi/2} \in U(1)_{B}$ generate a $\mathbb{Z}_{4}$ subgroup of $U(1)_{B}$.  The action of $J$ and $\omega$ on $A_{\mu}, q_{a}$ is
\be
A_{\mu} \rightarrow J A_{\mu} J^{T}, \;\; q_{a} \rightarrow -\omega J q_{a},
\ee
generating a $\mathbb{Z}_{2}$ subgroup of $SO(2N_{c})\times U(1)_{B}$. 

$A_{\mu}$ can be written in $N_{c}\times N_{c}$ blocks as
\begin{align}
A_\mu
=
i\left(
\begin{array}{cc}
A_\mu^A+B_\mu^A & C_\mu^A-D_\mu^S\\
C_\mu^A+D_\mu^S & A_\mu^A-B_\mu^A
\end{array}
\right),
\end{align}
where fields with an `$A$' (`$S$') superscript are antisymmetric (symmetric) matrices.  Under the $\mathbb{Z}_{2}$ symmetry, $A_\mu^A, D_\mu^S$ are even while $B_\mu^A, C_\mu^A$ are odd,  
so the orbifold projection sets $B_{\mu}^{A} = C_{\mu}^{A} = 0$.  So
\begin{align}
A_{\mu}^{proj}
=
i\left(
\begin{array}{cc}
A_{\mu}^A  & -D_\mu^S\\
D_\mu^S & A_\mu^A
\end{array}
\right).
\end{align} 
If one defines a unitary matrix
\be
P = \frac{1}{\sqrt{2}}\left(
\begin{array}{cc}
1_{N_{c}} & i 1_{N_{c}} \\
1_{N_{c}} & -i 1_{N_{c}}
\end{array} 
\right),
\ee
then
\be
P A_{\mu}^{proj} P^{-1} =
  \left(
\begin{array}{cc}
-\mathcal{A}_{\mu}^{T} & 0\\
0 & \mathcal{A}_{\mu}
\end{array} 
\right),
\ee
where $\mathcal{A}_{\mu} \equiv D_{\mu}^{S} + i A^{A}_{\mu}$ is a $U(N_{c})$ gauge field.  However, the difference between $U(N_{c})$ and $SU(N_{c})$ is a $1/N_{c}^{2}$ correction. The gauge part of the action of the orbifold-projected parent theory is thus simply
\be
\mathcal{L^{\mathrm{gauge}, \mathrm{proj}}} = \frac{2}{4g_{SO}^{2}} \Tr \mathcal{F}_{\mu \nu}\mathcal{F}^{\mu \nu}.
\ee 
where $\mathcal{F}_{\mu\nu}$ is the $SU(N_{c})$ field strength. 

Now consider the effect of the orbifold on $q_{a}$.  Writing $(\lambda^{+}_{a}, \lambda^{-}_{a})^{T} = (Pq_{a})^T$, the action of the $\mathbb{Z}_{2}$ symmetry is just  $(\lambda^{+}_{a}, \lambda^{-}_{a})^{T} \rightarrow (- \lambda^{+}_{a}, \lambda^{-}_{a})^{T}$. The projection consists of setting $\lambda^{+}_{a} = 0$.

The action of the daughter theory is the action of the parent theory after the projection, with a rescaled coupling constant $g_{SU} = g_{SO}$~\cite{Bershadsky:1998cb}
\be
\mathcal{L} = \frac{1}{4 g_{SU}^{2} } \Tr \mathcal{F}_{\mu \nu}^2
+ 
\sum_{a=1}^{N_f}
\bar{\psi}^{a}\left( \gamma^{\mu} {\cal D}_{\mu} + m_q + \mu_B\gamma^4\right)\psi_{a}
\ee
where $\mathcal{F}_{\mu\nu}$ is the field strength of the $SU(N_{c})$ gauge field $\mathcal{A}_{\mu} = D^{S}_{\mu} + i A^{A}_{\mu}$, $\psi_{a} = \lambda^{-}_{a}$, and ${\cal D}_{\mu} = \partial_{\mu} + i \mathcal{A}_{\mu}$.  This is an $SU(N_{c})$ gauge theory with $N_{f}$ flavors of fundamental Dirac fermions.   So the orbifold projection relates $SO(2N_{c})$ gauge theory to large $N_{c}$ QCD.  

{\it Neutral sector.}
The claim of orbifold equivalence is that the connected correlators of neutral operators in orbifold-equivalent parent and daughter theories will agree at large $N_{c}$.  We define neutral operators to be those that are invariant under the projection symmetry.  Color-singlet gluonic operators in the $SO(2N_{c})$ theory are neutral, and are mapped to $C$-even gluonic operators in $SU(N_c)$ theory by the projection.
\begin{table}[tdp]
\begin{center}
\begin{tabular}{|c|c|} \hline 
$SO(2N_{c})$ theory & $SU(N_{c})$ theory \\ \hline
$\bar{q}_{a} q_{b}$   & $\bar{\psi}_{a} \psi_{b}$ \\ \hline
$\bar{q}_{a} \gamma^{\mu} q_{b}$ & $\bar{\psi}_{a} \gamma^{\mu} \psi_{b}$ \\ \hline
$\bar{q}_{a} \gamma^{\mu} \partial_{\mu} q_{b}$ & $\bar{\psi}_{a} \gamma^{\mu} \partial_{\mu} \psi_{b}$ \\
\hline
\end{tabular} \caption{Examples of fermion bilinears whose correlation functions match between the two theories at large $N_{c}$. }
\end{center}
\label{table:Bilinears}
\end{table}

For fermionic observables, things are more subtle.  Some examples of fermion bilinears that survive the projection are given in Table~$\mathrm{I}$.  An example of a bilinear that does not survive the projection is $q^{T}_a C \gamma^{5} q_b$, the baryonic pion operator; the same is true for all of the baryonic mesons.  This is because all such operators have charge $-1$ under the $\mathbb{Z}_{2}$ projection symmetry.  Thus correlation functions involving baryonic mesons have no counterparts in large $N_{c}$ QCD.    Note that this implies that the counting of neutral NG bosons matches in the two theories~\cite{Kovtun:2005kh}, which is an important sanity check on the large $N_{c}$ equivalence.

{\it Validity of the equivalence.}
Given an orbifold projection between two theories, the correlation functions of neutral operators will agree in perturbation theory\cite{Bershadsky:1998cb}.  For an orbifold equivalence to hold nonperturbatively, the symmetries used in the projection must not be spontaneously broken \cite{Kovtun:2003hr,Kovtun:2004bz}.  While the proofs of Refs.~\cite{Kovtun:2003hr,Kovtun:2004bz} need be generalized to apply to projections involving fundamental fermions\cite{Cherman:2011mh,Hanada:2011ju}\footnote{The analysis may be especially subtle for the Veneziano large $N_{c}$ limit~\cite{Veneziano:1976wm}.}, it is natural to conjecture that the symmetry realization condition remains the key to the nonperturbative validity of the equivalence.  

In our case, $U(1)_{B} \rightarrow \mathbb{Z}_{2}$ when $\mu_{B} \geq m_{\pi}/2$ due to baryonic pion condensation.  This breaks the projection symmetry, destroying the equivalence.  All is not lost, however.  One can add a deformation term to the $SO(2N_{c})$ theory to prevent baryonic pion condensation and protect $U(1)_{B}$.  The simplest choice  is
\be
\mathcal{L}_{SO} \rightarrow \mathcal{L}_{SO} +  \frac{c^{2}}{\Lambda^{2}} \sum_{a,b}  S^{\dag}_{ab} S_{a b}
\ee 
where $\Lambda \sim \Lambda_{QCD}$.  The deformed theory should be viewed as an effective field theory, defined with an implicit (lattice) cutoff, since the deformation is an irrelevant operator.  Large $N_{c}$ factorization implies that the deformed system would pay an energy cost $\mathcal{O}(c^{2})$ for the formation of a baryonic pion condensate. Thus provided $c$ is large enough, the deformation prevents baryonic pion condensation, saving the validity of the orbifold equivalence between the deformed $SO(2N_{c})$ gauge theory and large $N_{c}$ QCD.  

By construction, deformation terms do not survive the orbifold projection to QCD, so the value of $c$ does not affect the correlation functions of neutral operators so long as the equivalence holds.  Much like the double-trace deformations used to prevent~\cite{Unsal:2008ch}  center-symmetry breaking in Eguchi-Kawai reduction~\cite{Eguchi:1982nm}, our deformation terms hide themselves once they do their job.

{\it The sign problem.}  
Consider the undeformed $SO(2N_{c})$ theory.   Then the Dirac operator $D = \dslash + m_{q}  + \mu_{B} \gamma_{4}$ satisfies $C\gamma^{5} D  (C\gamma^{5})^{-1}= D^{\ast}$.  If the lattice form of the Dirac operator also has this symmetry, then if $\varphi$ satisfies $D\varphi=\lambda\varphi$, $D(\gamma^5 C^{-1}\varphi^\ast)=\lambda^\ast(\gamma^5 C^{-1}\varphi^\ast)$, and $\varphi, \gamma^5 C^{-1}\varphi^{*}$ are orthogonal~\cite{Hands:2000ei}. So eigenvalues form pairs $(\lambda,\lambda^\ast)$, and hence $\det(D) \ge 0$, even at $\mu_{B}\neq 0$.

For simulations of the deformed theory, the action must be made quadratic in $q$, which can be arranged by `integrating in' auxiliary fields.  For general values of $m_{q},\mu_{B}$, the $C\gamma^{5} D (C\gamma^{5})^{-1}  = D^{\ast}$ symmetry of the Dirac operator is crucial for avoiding the sign problem, but deformations generically break it.  There are a variety of deformations that prevent baryonic pion condensation, and several ways of introducing auxiliary fields.  A deformation which turns out to have a simple effect on the deformed theory\cite{Cherman:2011mh} and can be implemented without a sign problem in the chiral limit is
\be
\mathcal{L}_{d} = \frac{c^{2}}{\Lambda_{\mathrm{QCD}}} \left(S^{\dag}{}^{ab} S_{a b} - P^{\dag}{}^{ab} P_{a b}  \right) 
\ee
where $P_{ab} = q_{a}^{T} C q_{b}$.  Using Fierz identities, this can be written as
\be
\mathcal{L}_{d} = \frac{c^{2}}{\Lambda_{\mathrm{QCD}}} \left[(\bar{q}^{i}_{a} q^{j}_{a})^{2} + (\bar{q}^{i}_{a} \gamma^{5} q^{j}_{a})^{2}+\frac{1}{2}(\bar{q}^{i}_{a} \gamma^{\mu \nu} q^{j}_{a})^{2}   \right] 
\ee
where there is an implied sum over the color labels $i,j$. We then introduce real auxiliary fields that couple to the flavor-singlet bilinears $\bar{q}_{i} \cdots q_{j}$.  This allows us to maintain a $C D C^{-1} = - D^{*}$ symmetry for any $c,\mu_{B}$ so long as  $m_{q}=0$, avoiding the sign problem.  While in practical lattice calculations  $m_{q} > 0$, the lack of a sign problem at $m_{q}=0$ implies that the phase-quenching approximation must become increasingly accurate as $m_{q}\to 0$ in this theory.

The sign-free deformation may seem peculiar, especially since $\mathcal{L}_{d}$ is not positive definite.  However, one can get a nonperturbative understanding of its effects using low-energy effective field theory.  The result of this analysis, which will be presented elsewhere\cite{Cherman:2011mh}, is very simple:  the deformation raises the mass of the baryonic pions, pushing their condensation point past $\mu_{B} = m_{\pi}/2$, saving the equivalence.  Once $\mu_{B} \sim \Lambda_{QCD}$, baryonic mesons with masses $\sim \Lambda_{QCD}$ might condense.  Whether this happens depends on which states in the deformed theory have the smallest mass per $U(1)_{B}$ charge, and will have to be resolved by lattice simulations.  It would be very interesting if the equivalence works through the nuclear matter transition.  

{\it More applications.}
The arguments so far also hold at finite temperature. In particular, the details of the chiral transition can be studied as long as the baryonic pion does not condense.  So even without the deformation, one can gain valuable insights into hot QCD.   Our framework also gives insights into the behavior of phase-quenched simulations.  When $N_{f}$ is even, one can do a projection of the SO theory by using a $\mathbb{Z}_{4}$ subgroup of the $U(1)_{I_{3}} \in SU(N_{f})_{V}$ ``isospin'' flavor symmetry instead of $U(1)_{B}$.  The daughter theory is again large $N_{c}$ QCD, but now $\mu_{B}$ in the parent is mapped to an isospin chemical potential $\mu_{I}$ in QCD.    The resulting daughter-daughter equivalence between QCD with $\mu_{B}\neq0$ and $\mu_{I}\neq0$ holds for $\mu<m_\pi/2$.    It has been previously noticed that phase-quenching QCD with $\mu_{B} < m_{\pi}/2$ seems to be a good approximation for some observables, and that the phase-quenched theory is just QCD with $\mu_{I} \neq 0$~\cite{Alford:1998sd}.   The daughter-daughter equivalence  guarantees that at large $N_c$ phase quenching is exact for observables with zero baryon and isospin charges, giving additional insights into the behavior of this approximation\footnote{The exactness of phase quenching for the chiral condensate in random matrix models can also be shown to follow from orbifold equivalence\cite{Hanada:2011ju}.}.  The equivalence can be extended to a holographic setup\cite{HanadaHoyosKarch},  and the coincidence of the phase diagram in the baryonic\cite{Mateos:2007vc} and isospin\cite{Erdmenger:2008yj} theories can be seen analytically.

{\it Outlook. } 
We have proposed a way to dodge the sign problem in the chiral limit of large $N_{c}$ QCD by working with a large-$N_{c}$ equivalent $SO(2N_{c})$  theory.  There are many directions for future work, some of which were mentioned above.  Of these, tests of the proposal on the lattice and EFT analysis of the IR physics of the deformed theory are perhaps the most urgent.   Finally, one might wonder if orbifold equivalence can allow one to dodge sign problems in other systems, for instance in SYM theories~\cite{Krauth:1998xh}.  This would make the Monte Carlo approach to the gauge-gravity duality (see, {\it e.g.} \cite{Hanada:2009ne}) much more tractable.   

{\it Acknowledgements.}  We thank O.~Aharony, T.~Azeyanagi, P.~Bedaque, E.~Berkowitz, M.~Buchoff, T.~Cohen, M.~Tezuka, M.~Unsal, L.~Yaffe, and N.~Yamamoto for very stimulating discussions.  We especially thank  B.~Tiburzi for sharing many crucial insights, particularly on the parity of the baryonic pions.   We also thank A.~Armoni and J.~C.~Myers for useful comments at an early stage.  A.~C. was supported by the U.S. DOE Grant No. DE-FG02-93ER-40762.

\bibliographystyle{apsrev4-1}
\bibliography{orbifoldingNoArxiv} 

\end{document}